\documentclass[lettersize,journal]{IEEEtran}

\usepackage{lettrine}
\usepackage[dvipsnames]{xcolor}
\usepackage{hyperref}
\newcommand\myshade{85}
\colorlet{mylinkcolor}{blue}
\colorlet{mycitecolor}{YellowOrange}
\colorlet{myurlcolor}{Aquamarine}
\hypersetup{
linkcolor  = mylinkcolor!\myshade!black,
citecolor  = mycitecolor!\myshade!black,
urlcolor   = myurlcolor!\myshade!black,
colorlinks = true,}
\newcommand{\eq}[1]{{\fontfamily{cmr}\selectfont\color{blue!70!black}#1}}
\usepackage{amsmath}
\usepackage{tikz}
\usetikzlibrary{arrows.meta, positioning, shapes.geometric, fit, backgrounds}
\tikzset{
box/.style = {draw, thick, rounded corners, align=center, fill=white, inner sep=3pt, minimum width=3.2cm},
arrow/.style = {thick, -{Latex[length=2.5mm,width=1.6mm]}},
small/.style = {font=\footnotesize}}

\usepackage{booktabs}
\usepackage{cite}
\usepackage{graphicx}
\usepackage{enumitem}
\usepackage{amssymb,mathtools}
\usepackage{siunitx}
\usepackage[ruled,vlined,linesnumbered]{algorithm2e}
\usepackage[dvipsnames]{xcolor}
\newcommand{\blue}[1]{#1}
\usepackage{xcolor}
\definecolor{slate}{RGB}{112, 128, 144}
\definecolor{modernslate}{HTML}{708090}
\definecolor{slatelight}{RGB}{240, 242, 245}
\definecolor{amber}{RGB}{255, 191, 0}
\definecolor{profamber}{RGB}{255, 174, 66}
\definecolor{indigo}{RGB}{75, 0, 130}
\definecolor{profindigo}{HTML}{3F51B5}
\definecolor{modernamber}{HTML}{FFBF00}

\begin{document}
	\title{{DAE-Aware Bayesian Inference for Joint Generator-Network Parameter Estimation}}
	
	\author{Abdallah Alalem Albustami*, Ahmad F. Taha, and Sankaran Mahadevan 
		\thanks{*Corresponding author. Authors are with the Civil Engineering Department at
			Vanderbilt University, Nashville, TN 37235, USA.
			(e-mails: \{abdallah.b.alalem.albustami, sankaran.mahadevan ahmad.taha\}@vanderbilt.edu).}
		}

    \markboth{In Press, \textit{Electric Power Systems Research}; To appear in the 24TH Power Systems Computation Conference (PSCC 2026), Cyprus}{}

	\maketitle

\begin{abstract}
This paper addresses the classic problem of parameter estimation (PE) in multimachine power system models. Such models are typically described by a set of nonlinear differential-algebraic equations (DAE), where generator physics and network power flow equations are coupled. DAE models are well established in classic power system textbooks, but parameter identification and estimation of generator inertia and damping together with network branch resistances and reactances for these models remain relatively underexplored. In contrast to prior approaches that rely on ODE approximations, this paper develops a joint Bayesian inference framework to perform PE of generator and network parameters while exploiting grid DAE models. It further combines physics-aware statistical modeling with computationally efficient posterior sampling to make joint Bayesian calibration practical. Results on the IEEE 9-bus system show accurate parameter recovery with well-behaved posterior uncertainty, while a short 39-bus study provides evidence that the framework remains effective on a materially larger joint-estimation problem. These results are obtained without requiring overly conservative priors.
\end{abstract}


\section{Introduction and Paper Contributions}
\lettrine[lraise=0.1, nindent=0em, slope=-.5em]{M}{odern}  power systems operate closer to stability limits as renewable integration increases and conventional synchronous generation decreases. This transition creates two problems: declining system inertia makes frequency dynamics faster while the dynamic parameters governing these phenomena become harder to specify. Inverter-based resources have variable or synthetic inertia that cannot be determined from nameplate data. Aging synchronous generators have inertias and dampings that drift from manufacturer specifications due to decades of operation and rewinding. The net result is that the dynamic models used for stability assessment increasingly diverge from physical reality.

Parameter estimation (PE) from measurement data offers a solution. Phasor measurement units (PMUs) provide synchronized voltage phasor-derived channels and frequency measurements at 30--1200 Hz across transmission networks and capture the electromechanical transients that reveal dynamic parameters. The pertinent question is how to perform PE and compute their uncertainties from PMU measurements in a way that respects the underlying physics and provides operators with actionable information about model confidence. Power systems PE has been studied extensively with methods falling virtually into four main categories as discussed next.

\noindent {\textbf{Model-based estimation.}} Classical approaches formulate PE as optimization using known physical models. Weighted least squares (WLS) methods fit transmission line parameters by minimizing measurement residuals with appropriate noise weighting \cite{mousavi2014parameter}. Total least squares (TLS) variants account for errors in both voltage and current measurements \cite{albuquerque2021estimation}. State estimation extensions treat unknown parameters as augmented state variables. 
Kalman filtering approaches augment the state vector with unknown parameters, recursively updating both states and parameters as measurements arrive. Extended Kalman filters (EKF) have been applied to estimate generator inertia, damping, and transient reactances from PMU measurements \cite{fan2013extended}. Unscented Kalman filters (UKF) address nonlinearity of grid models through deterministic sampling and have been validated on real transmission system data \cite{valverde2011unscented}. Frequency-domain identification methods extract parameters from disturbance response using \textit{Prony} analysis or \textit{ARMAX} models fitted to ambient measurements \cite{liu2015armax}. 

\noindent {\textbf{Optimization-based methods.}} Deterministic nonlinear least squares uses gradient-based algorithms (such as Gauss-Newton or Levenberg-Marquardt) to minimize measurement residuals. Metaheuristics such as genetic algorithms, particle swarm optimization, and differential evolution can escape local minima but require many model evaluations \cite{hutchison2010synchronous, hutchison2015non}.

\noindent {\textbf{Bayesian inference.}} Bayesian methods treat parameters as random variables with prior distributions, updating beliefs using measurement data to obtain posterior distributions that quantify uncertainty. This provides credible intervals and confidence levels, not just point estimates. Petra \textit{et al.} applied Bayesian inference with adjoint-based sensitivity analysis to estimate generator inertias, demonstrating how posterior variances reflect measurement quality \cite{petra2016bayesian}. Standard Metropolis-Hastings MCMC mixes slowly in high dimensions. Recent advances use affine-invariant ensemble samplers that deploy multiple parallel walkers achieving faster convergence for multi-generator inertia estimation \cite{zhong2025bayesian}. Bayesian methods are attractive because they provide principled uncertainty quantification, but their application to power system PE remains challenging. In much of the literature, Bayesian inference has been used mainly for a limited subset of dynamic parameters, especially generator inertias or related machine parameters, rather than fully joint estimation of dynamic and network quantities \cite{petra2016bayesian}. A second challenge is computational scalability, sampling-based Bayesian inference can require many repeated simulations of large nonlinear DAE models, which quickly becomes expensive as system size and parameter dimension grow \cite{petra2016bayesian, nagi2021bayesian}. While some recent methods improve efficiency, the broader issues of strong parameter coupling, model mismatch, and scalability in large power networks remain active challenges \cite{zhong2025bayesian, nagi2021bayesian}.

\noindent {\textbf{Machine learning methods.}} Data-driven approaches bypass explicit physical models. Neural networks have been trained to predict inertia from frequency time-series using convolutional architectures \cite{linaro2023continuous}. Graph neural networks exploit system connectivity to estimate generator inertias from PMU measurements \cite{albeladi2025beyond, tuo2023machine}. These methods can capture complex nonlinear relationships but require large training datasets and provide limited interpretability. Physics-informed neural networks (PINNs) embed governing equations into training, treating unknown parameters as learnable variables that must satisfy both data and differential equations \cite{lakshminarayana2022application}. While these methods are flexible and can model nonlinear dynamics, they typically require large amounts of training data and are often tied to a fixed network topology. For example, standard neural network and graph neural network models generally need retraining when network parameters change. PINNs include physical constraints during training, but these are usually imposed as soft penalties rather than exact feasibility conditions. Moreover, these approaches do not generally provide calibrated uncertainty estimates in a natural way.

\noindent {\textbf{Literature gaps and paper contributions.}}  The gap in existing work is simply that no prior method jointly estimates generator dynamic parameters and network parameters from the same PMU dataset while accounting for their coupling through the DAE structure. While many studies address single-machine parameter estimation (such as single-machine infinite-bus models for inertia/damping) and multi-machine estimation (such as system-wide inertia identification) using aforementioned approaches, all treat either generator parameters or network parameters as known.

PE for DAE grid models matters because power systems are not modular, one cannot separate generator dynamics from network dynamics and estimate them independently. The swing equation depends on electrical power,  which depends on network admittances present in the power flow model. If generator parameters are estimated with wrong network parameters, the estimated inertias and dampings then compensate for network errors. The resulting model might fit historical data but will fail when network topology or loading changes. Similarly, estimating network parameters while treating generator dynamics incorrectly leads to biased admittance estimates. This coupling is not merely a secondary modeling detail, it directly affects how generator and network parameters are inferred from the same transient measurements.

\textit{PE for DAE grid models is hard. } For example, considering the IEEE 9-bus test system with 3 generators, the unknown vector contains 21 parameters, and many combinations produce nearly indistinguishable transient responses. The resulting inverse problem is ill-conditioned, so standard optimization faces flat objectives and local minima, while standard MCMC mixes slowly and requires repeated stiff DAE solves.

The contribution of this work is a DAE-aware Bayesian calibration framework that jointly estimates generator inertias/dampings and network branch resistances/reactances while restricting inference to parameter values that admit feasible power-flow initialization. The method \textit{(a)} exploits transient time-scale separation through segmented likelihood, \textit{(b)} introduces model-discrepancy inflation, \textit{(c)} uses log-normal standardized reparameterization with physics-informed priors, and \textit{(d)} accelerates sampling through blocked proposals and multifidelity delayed acceptance. We quantify generator--network coupling and demonstrate accurate recovery with uncertainty quantification on IEEE 9-bus and 39-bus systems.

\noindent \textbf{What this paper is \textit{not} and paper organization.} This paper is a proof of concept for a hard power-system inverse problem. It adapts established Bayesian/MCMC concepts to power system DAEs and demonstrates that reliable power system PE is possible without conservative priors. We do not claim global identifiability or general convergence guarantees beyond stated assumptions. Scaling up the proposed method is a difficult problem, one we do not pursue herein (which remains an issue with all other studies in the literature). 

The remainder of this paper is organized as follows. Section \ref{sec:prelim} includes preliminaries and the power system DAE model. Section \ref{subsec:identifiability} presents identifiability analysis and problem formulation. Section \ref{sec:bayes} develops the DAE-aware Bayesian Inference Framework. Section \ref{sec:pf_mcmc} introduces the multifidelity delayed-acceptance MCMC approach. Section \ref{sec:case_studies} presents case studies, and Section \ref{sec:conc} concludes the paper.

\section{Preliminaries: Power System Model} \label{sec:prelim}
We consider a fourth-order model with $n_N$ buses belonging to set $\mathcal{N}$ and $n_G$ synchronous machines belonging to set $\mathcal{G} \subset \mathcal{N}$; the framework introduced in this paper can be generalized for higher (or lower) order grid models. The dynamic generator state is defined by vector $\boldsymbol{x}_d(t)\in\mathbb{R}^{n_d}$ which includes rotor angle $\delta_i(t)$, electrical speed $\omega_i(t)$, transient internal EMF $E'_i(t)$, and mechanical torque $T_{m,i}(t)$ for all $i \in \mathcal{G}$. The algebraic state vector is depicted by $\boldsymbol{x}_a(t)\in\mathbb{R}^{n_a}$ which includes active/reactive injections $P_{G,i}(t),Q_{G,i}(t)$ for $i\in\mathcal{G}$ together with bus voltage magnitudes $V_j(t)$ and bus angles $\theta_j(t)$ for $j\in\mathcal{N}$, so $n_a=2n_G+2n_N$.

The set of parameters to calibrate consists of generator parameters (inertias $\boldsymbol{M}$ and damping coefficients $\boldsymbol{D}$) and the network parameters (series branch resistances $\boldsymbol{r}$ and reactances $\boldsymbol{x}$ for branches with nonzero values). All other branch data, including line charging susceptances and any fixed topology data, are treated as known and held at their nominal values. The bus admittance matrix $\boldsymbol{Y}_{\mathrm{bus}}=\boldsymbol{G}+j\boldsymbol{B}$ is assembled deterministically from these branch parameters (see~\eqref{eq:Y-assembly}), so $\boldsymbol{G}$ and $\boldsymbol{B}$ are derived quantities rather than free parameters. All unknowns are collected in vector $\boldsymbol{\theta}$
\begin{align}
\boldsymbol{\theta} \;=\; \big[\,\boldsymbol{M}^\top\ \boldsymbol{D}^\top\ \boldsymbol{r}^\top\ \boldsymbol{x}^\top\big]^\top 
\ \in\ \mathbb{R}^{n_\theta}, \label{eq:theta}
\end{align}
with $\boldsymbol{M},\boldsymbol{D}\in\mathbb{R}^{n_G}_{>0}$, $\boldsymbol{r}\in\mathbb{R}^{n_r}_{>0}$ collecting the nonzero branch resistances, and $\boldsymbol{x}\in\mathbb{R}^{n_x}_{>0}$ collecting the nonzero branch reactances. The dimension is $n_\theta \;=\; 2n_G \;+\; n_r \;+\; n_x.$
Considering that $\boldsymbol{x}(t)$ stacks both $\boldsymbol{x}_d(t)$ and $\boldsymbol{x}_a(t)$, and assuming PMUs measure selected rectangular voltage components together with generator frequency deviations through vector $\boldsymbol{y}(t_k) \in \mathbb{R}^{p}$ at discrete time-steps $t_k$, the multi-machine power grid model can be written as a semi-explicit index-1 DAE \cite{Kundur2017}: 
\begin{subequations}\label{eq:dae-compact-f} 
\begin{align}
\boldsymbol{E}\,\dot{\boldsymbol{x}}(t)&= \boldsymbol{f}\!\left(\boldsymbol{x},\boldsymbol{u};\boldsymbol{\theta}\right), \;\; \boldsymbol{0} = \boldsymbol{g}\!\left(\boldsymbol{x},\boldsymbol{d};\boldsymbol{\theta}\right), \label{eq:dae-compact} \\
\boldsymbol{y}(t_k) &= \boldsymbol{h}_{\mathrm{meas}}\!\left(\boldsymbol{x}_d(t_k;\boldsymbol{\theta}),\boldsymbol{x}_a(t_k;\boldsymbol{\theta})\right) \;+\; \boldsymbol{\varepsilon}(t_k),  \label{eq:meas}
\end{align}
\end{subequations}
with inputs/disturbances $\boldsymbol{u}(t)$ and loads $\boldsymbol{d}(t)$. Singular matrix $\boldsymbol{E}$ encodes the algebraic power flow constraints through blocks of zeroes. Quantity $\boldsymbol{\varepsilon}(t_k)\sim\mathcal{N}(\boldsymbol{0},\boldsymbol{R})$ defines measurement noise, and $\boldsymbol{h}_{\mathrm{meas}}(\cdot)$ denotes the measurement map. The voltage channels are built from algebraic voltage states, while the frequency channels depend on generator speed states.
The full dynamic state $\boldsymbol{x}_d$ is not directly observed, only selected generator frequency deviations are available.
The function $\boldsymbol{f}(\cdot)$ represents the differential equations for generator dynamics, and $\boldsymbol{g}(\cdot)$ represents the power flow and power balance constraints. The actual models are omitted for brevity and can be found in \cite{nadeem2022dynamic}. The proposed framework is generalizable to any semi-explicit index-1 power system DAE, replacing the fourth-order machine with higher-order synchronous or inverter models only changes functions $\boldsymbol{f}$ and $\boldsymbol{g}$.

The bus admittance matrix is assembled from branch parameters via standard power system conventions~\cite{bergen2009power}:
\begin{subequations}
\label{eq:Y-assembly}
\begin{flalign}
&G_{ij} =-\,g^{\mathrm{s}}_{(i,j)}, \;
B_{ij} =-\,b^{\mathrm{s}}_{(i,j)}, \;
 i\neq j,\ (i,j)\in\mathcal{E}, && \label{eq:Y-off}\\
&G_{ii} = \hspace{-0.3cm}\sum_{j:(i,j)\in\mathcal{E}}\hspace{-0.3cm} g^{\mathrm{s}}_{(i,j)} + g_i^{\mathrm{sh}}, \;
B_{ii} = \hspace{-0.3cm}\sum_{j:(i,j)\in\mathcal{E}} \hspace{-0.3cm}b^{\mathrm{s}}_{(i,j)} + b_i^{\mathrm{sh}},&&  \label{eq:Y-diag}
\end{flalign}
\end{subequations}
where the series admittances $(g^{\mathrm{s}}_{(i,j)}, b^{\mathrm{s}}_{(i,j)})$ are computed from the branch resistances and reactances via $g^{\mathrm{s}} = r/(r^2+x^2)$ and $b^{\mathrm{s}} = -x/(r^2+x^2)$, and $g_i^{\mathrm{sh}}, b_i^{\mathrm{sh}}$ are known bus shunt elements. This construction automatically enforces symmetry and sign patterns from the physical branch parameters.

Given $\boldsymbol{\theta}$ and a pre-disturbance operating point $(\boldsymbol{u}_0,\boldsymbol{d}_0)$, the initial state $(\boldsymbol{x}_d^\star,\boldsymbol{x}_a^\star)$ must satisfy
\begin{align}
\boldsymbol{f}\!\left(\boldsymbol{x}^\star,\boldsymbol{u}_0;\boldsymbol{\theta}\right)=\boldsymbol{0},\qquad
\boldsymbol{g}\!\left(\boldsymbol{x}^\star,\boldsymbol{d}_0;\boldsymbol{\theta}\right)=\boldsymbol{0}. \label{eq:pfcons}
\end{align}
The set of all such equilibria forms the power-flow (PF) manifold $\mathcal{M}_{\text{PF}}(\boldsymbol{\theta})$. To do parameter estimation through Bayesian inference, we must repeatedly evaluate the forward model: given a candidate $\boldsymbol{\theta}$, we need to solve~\eqref{eq:pfcons} for a consistent initial condition $(\boldsymbol{x}^\star_d,\boldsymbol{x}^\star_a)\in\mathcal{M}_{\text{PF}}(\boldsymbol{\theta})$, then integrate the DAE~\eqref{eq:dae-compact} from $t=0$ to $T$ to obtain predicted measurements. These predictions are compared against the actual PMU measurements $\boldsymbol{y}(t_k)$ to assess how well $\boldsymbol{\theta}$ explains the observed data. This forward solve must be repeated for each parameter candidate during the estimation process.

\section{Local Identifiability Diagnostics and Standardized Parameterization}
\label{subsec:identifiability}
Prior to estimating parameters, we must address a fundamental question: which parameters can be recovered from PMU measurements, and how do generator and network parameters interact through the DAE coupling? Classical identifiability analysis for ODEs~\cite{richmond1974identifiability} does not directly apply to DAE systems where algebraic constraints couple parameter groups. Several papers discuss identifiability analysis for DAEs but not in the context of power system parameter estimation with mixed voltage and frequency output channels \cite{chen2013new, chen2014extended, landgraf2022analysis}. This section develops this DAE-aware identifiability framework.

Prior parameter estimation work in power systems treats generator dynamics and network parameters separately, assuming the other is known. This decoupling ignores the fact that inertia $M_i$ affects frequency dynamics which depend on electrical power $P_e$, which in turn depends on network admittances through the power flow equations. Estimating generators with incorrect network parameters yields biased inertia estimates that compensate for network errors. The identifiability analysis presented here quantifies this coupling explicitly via the DAE structure, revealing which parameter combinations are jointly identifiable.
The outcome of this section is: \textit{(i)} a local Gauss-Newton curvature diagnostic, \textit{(ii)} a \emph{co-identifiability map}, and \textit{(iii)} a physical branch-parameter reparameterization for efficient MCMC sampling.

\subsection{Forward DAE Sensitivities}
\label{subsec:forward_sens}

To assess how measurements $\boldsymbol{y}(t_k)$ change with parameters $\boldsymbol{\theta}$, one may define sensitivity matrices $\boldsymbol{S}_d(t):=\partial\boldsymbol{x}_d(t)/\partial\boldsymbol{\theta}\in\mathbb{R}^{n_d\times n_\theta}$ and $\boldsymbol{S}_a(t):=\partial\boldsymbol{x}_a(t)/\partial\boldsymbol{\theta}\in\mathbb{R}^{n_a\times n_\theta}$. These quantify how small parameter perturbations propagate through the DAE to affect states. Differentiating~\eqref{eq:dae-compact} with respect to $\boldsymbol{\theta}$ yields the coupled variational system:
\begin{subequations}\label{eq:sens-dae}
\begin{align}
\dot{\boldsymbol{S}}_d(t) &= 
\boldsymbol{f}_{x_d}(t)\,\boldsymbol{S}_d(t) + \boldsymbol{f}_{x_a}(t)\,\boldsymbol{S}_a(t) + \boldsymbol{f}_{\theta}(t), 
\label{eq:sd-eq}\\
\boldsymbol{0} &= 
\boldsymbol{g}_{x_d}(t)\,\boldsymbol{S}_d(t) + \boldsymbol{g}_{x_a}(t)\,\boldsymbol{S}_a(t) + \boldsymbol{g}_{\theta}(t),
\label{eq:sa-constraint}
\end{align}
\end{subequations}
where all Jacobians are evaluated along the trajectory $(\boldsymbol{x}_d(t;\boldsymbol{\theta}),\boldsymbol{x}_a(t;\boldsymbol{\theta}),\boldsymbol{u}(t))$. Specifically, $\boldsymbol{f}_{x_d}(t):=\partial\boldsymbol{f}/\partial\boldsymbol{x}_d|_{(\boldsymbol{x}(t),\boldsymbol{u}(t);\boldsymbol{\theta})}$, and similarly for other Jacobians. Since~\eqref{eq:sa-constraint} is algebraic, at each time $t$ we solve for $\boldsymbol{S}_a(t)$:
\begin{align}
\boldsymbol{S}_a(t) = -\boldsymbol{g}_{x_a}(t)^{-1}\big(\boldsymbol{g}_{x_d}(t)\,\boldsymbol{S}_d(t) + \boldsymbol{g}_{\theta}(t)\big),
\label{eq:sa-explicit}
\end{align}
and substitute into~\eqref{eq:sd-eq} to obtain an ODE for $\boldsymbol{S}_d(t)$ alone:
\begin{equation}
\begin{aligned}
\dot{\boldsymbol{S}}_d(t) = 
\underbrace{\big(\boldsymbol{f}_{x_d}(t)-\boldsymbol{f}_{x_a}(t)\boldsymbol{g}_{x_a}(t)^{-1}\boldsymbol{g}_{x_d}(t)\big)}_{\boldsymbol{A}_s(t)}\,\boldsymbol{S}_d(t)
+\\
\underbrace{\big(\boldsymbol{f}_{\theta}(t)-\boldsymbol{f}_{x_a}(t)\boldsymbol{g}_{x_a}(t)^{-1}\boldsymbol{g}_{\theta}(t)\big)}_{\boldsymbol{B}_s(t)}.
\label{eq:sd-ode}
\end{aligned}
\end{equation}
The time-varying matrices $\boldsymbol{A}_s(t)$ and $\boldsymbol{B}_s(t)$ encode how parameter sensitivities evolve. The Schur complement terms $\boldsymbol{g}_{x_a}^{-1}\boldsymbol{g}_{x_d}$ and $\boldsymbol{g}_{x_a}^{-1}\boldsymbol{g}_{\theta}$ route the influence of generator parameters $(\boldsymbol{M},\boldsymbol{D})$ and network parameters $(\boldsymbol{r},\boldsymbol{x})$ through the algebraic constraints into the dynamic sensitivities $\boldsymbol{S}_d(t)$. This is where DAE coupling enters the identifiability analysis.

Initial sensitivities at the pre-disturbance equilibrium $(\boldsymbol{x}_d^\star,\boldsymbol{x}_a^\star)$ follow from differentiating the equilibrium conditions $\boldsymbol{f}=\boldsymbol{0}$ and $\boldsymbol{g}=\boldsymbol{0}$ in~\eqref{eq:pfcons}:
\begin{align}
\begin{bmatrix}
\boldsymbol{f}_{x_d} & \boldsymbol{f}_{x_a} \\
\boldsymbol{g}_{x_d} & \boldsymbol{g}_{x_a}
\end{bmatrix}_{\!(\boldsymbol{x}^\star,\boldsymbol{\theta})}
\begin{bmatrix}
\boldsymbol{S}_d(0)\\[2pt] \boldsymbol{S}_a(0)
\end{bmatrix}
=
-
\begin{bmatrix}
\boldsymbol{f}_{\theta} \\[2pt] \boldsymbol{g}_{\theta}
\end{bmatrix}_{\!(\boldsymbol{x}^\star,\boldsymbol{\theta})}.
\label{eq:init-sens}
\end{align}

Equations~\eqref{eq:sa-explicit}--\eqref{eq:sd-ode} describe the continuous-model sensitivity structure and clarify how generator and network parameters couple through the DAE constraints. At a PMU sample time $t_k$, the corresponding measurement sensitivity satisfies the chain-rule relation
\begin{align*}
\boldsymbol{J}_k := \frac{\partial \boldsymbol{y}(t_k)}{\partial \boldsymbol{\theta}}
=
\boldsymbol{h}_{x_d}(t_k)\,\boldsymbol{S}_d(t_k)+\boldsymbol{h}_{x_a}(t_k)\,\boldsymbol{S}_a(t_k)
\in \mathbb{R}^{p\times n_\theta},
\end{align*}
where $\boldsymbol{h}_{x_d}(t_k):=\partial \boldsymbol{h}_{\mathrm{meas}}/\partial \boldsymbol{x}_d|_{t_k}$ and $\boldsymbol{h}_{x_a}(t_k):=\partial \boldsymbol{h}_{\mathrm{meas}}/\partial \boldsymbol{x}_a|_{t_k}$.
In the present implementation, however, the local curvature diagnostic used for initialization and proposal preconditioning is constructed from finite-difference Jacobians of the stacked residual map at the reference point, rather than from explicit integration of the sensitivity system above.

\subsection{DAE Coupling and Co-Identifiability Map}
\label{subsec:coident}

The physical parameter vector used in estimation consists of generator inertias and dampings together with the uncertain series branch resistances $\boldsymbol{r}$ and reactances $\boldsymbol{x}$ from the network topology. The bus admittance matrix $\boldsymbol{Y}_{\mathrm{bus}}=\boldsymbol{G}+j\boldsymbol{B}$ is assembled from these branch parameters via~\eqref{eq:Y-assembly}, so that $\boldsymbol{G}$ and $\boldsymbol{B}$ are \emph{derived} quantities rather than free parameters. This parameterization exploits the known network topology to dramatically reduce the dimension of the estimation problem: only branches with nonzero resistance or reactance appear in the parameter vector.
At a chosen reference point in standardized coordinates, we define the local Gauss--Newton curvature matrix by
$\boldsymbol{H}:=\boldsymbol{J}_{\eta}^{\top}\boldsymbol{W}\boldsymbol{J}_{\eta},$
where $\boldsymbol{J}_{\eta}$ is the stacked measurement Jacobian with respect to $\boldsymbol{\eta}$ and $\boldsymbol{W}$ is the diagonal weighting induced by the effective noise model and time-segment weights. Thus, $\boldsymbol{H}$ is a local curvature diagnostic, not a global identifiability certificate.
The parameter vector ordering $\boldsymbol{\theta}=[\boldsymbol{M}^\top,\boldsymbol{D}^\top,\boldsymbol{r}^\top,\boldsymbol{x}^\top]^\top$ induces a natural block structure in the local curvature matrix $\boldsymbol{H}$. We partition:
\begin{align*} \small
\boldsymbol{H} =
\begin{bmatrix}
\boldsymbol{H}_{MM} & \boldsymbol{H}_{MD} & \boldsymbol{H}_{Mr} & \boldsymbol{H}_{Mx} \\
\boldsymbol{H}_{DM} & \boldsymbol{H}_{DD} & \boldsymbol{H}_{Dr} & \boldsymbol{H}_{Dx} \\
\boldsymbol{H}_{rM} & \boldsymbol{H}_{rD} & \boldsymbol{H}_{rr} & \boldsymbol{H}_{rx} \\
\boldsymbol{H}_{xM} & \boldsymbol{H}_{xD} & \boldsymbol{H}_{xr} & \boldsymbol{H}_{xx}
\end{bmatrix},
\quad
\boldsymbol{H}_{XY}\in\mathbb{R}^{n_X\times n_Y},
\end{align*}
where $\boldsymbol{H}_{XY}$ is the cross-block curvature between groups $X$ and $Y$, and $n_X$ is the number of parameters in group $X$. 

The DAE structure enters twice: \textit{(i)} through Schur terms in~\eqref{eq:sd-ode}, which route network parameter influences into generator dynamic sensitivities, and \textit{(ii)} through~\eqref{eq:sa-explicit} which maps generator sensitivities back to algebraic measurements via $\boldsymbol{g}_{x_a}^{-1}$. As a result, off-diagonal blocks $\boldsymbol{H}_{Mr}, \boldsymbol{H}_{Mx}, \boldsymbol{H}_{Dr}, \boldsymbol{H}_{Dx}$ are generally nonzero, generator and network parameters are coupled through the measurements.

To summarize coupling strength across parameter groups, we define the normalized co-identifiability index:
\begin{align} \small
I_{XY} := \frac{\|\boldsymbol{H}_{XY}\|_F}{\sqrt{\|\boldsymbol{H}_{XX}\|_F\,\|\boldsymbol{H}_{YY}\|_F}}, \quad X,Y\in\{M,D,r,x\}. 
\label{eq:coid}
\end{align}
This normalizes the cross-block Frobenius norm by the geometric mean of diagonal block norms. Values near 1 indicate strong local coupling between parameter groups, while values near 0 suggest weak local coupling at the chosen reference point. The $4\times 4$ matrix $\{I_{XY}\}$ is the \emph{co-identifiability map}.To illustrate, a high $I_{Mr}$ value means inertia and branch resistances are strongly coupled in their local effect on measurements, so estimating one while fixing the other is more likely to introduce bias. This map serves as a local diagnostic for whether joint estimation is likely to be beneficial.

\subsection{Log-Normal Reparameterization}
\label{subsec:reduced}

All physical parameters---inertias, dampings, resistances, and reactances---are strictly positive and subject to multiplicative uncertainty (aging, temperature, modeling error). We therefore work in log-space throughout, defining latent coordinates $\boldsymbol{\lambda}=\log\boldsymbol{\theta}\in\mathbb{R}^{n_\theta}$. This ensures positivity by construction and maps the multiplicative prior structure to additive Gaussian perturbations, which is natural for MCMC proposals. The prior in latent space is $\lambda_i\sim\mathcal{N}(\log\theta_i^{\mathrm{nom}},\sigma_i^2)$ where $\theta_i^{\mathrm{nom}}$ is the nominal value and $\sigma_i$ controls the prior width.

To further standardize the sampling geometry, we define unit-variance coordinates $\boldsymbol{\eta}\in\mathbb{R}^{n_\theta}$ via
\begin{align}
\boldsymbol{\eta} = (\boldsymbol{\lambda} - \boldsymbol{\mu}_\lambda) \oslash \boldsymbol{\sigma}_\lambda, \qquad \boldsymbol{\lambda} = \boldsymbol{\mu}_\lambda + \boldsymbol{\sigma}_\lambda \odot \boldsymbol{\eta},
\label{eq:eta-param}
\end{align}
where $\boldsymbol{\mu}_\lambda=\log\boldsymbol{\theta}^{\mathrm{nom}}$ and $\boldsymbol{\sigma}_\lambda$ collects the per-parameter prior standard deviations. In $\boldsymbol{\eta}$-space, the Gaussian part of the prior simplifies to $p(\boldsymbol{\eta})=\mathcal{N}(\boldsymbol{0},\boldsymbol{I})$, so that within the admissible box the log-prior is $-\tfrac{1}{2}\|\boldsymbol{\eta}\|^2$, up to a constant. All MCMC sampling operates in $\boldsymbol{\eta}$-coordinates. Whenever the forward DAE model must be evaluated, the physical parameters are recovered via $\boldsymbol{\theta}=\exp(\boldsymbol{\mu}_\lambda+\boldsymbol{\sigma}_\lambda\odot\boldsymbol{\eta})$.

This reparameterization offers three advantages: \textit{(i)} it preserves all physical parameter directions so that weakly identifiable but physically meaningful parameters (e.g., small branch resistances) are still estimated with proper uncertainty rather than projected away; \textit{(ii)} it avoids computing and storing the full FIM eigenbasis, which becomes expensive for large systems; and \textit{(iii)} the unit-variance geometry makes the Hessian of the log-prior exactly $\boldsymbol{I}$, improving the condition number of the posterior and accelerating MCMC mixing. Hard box constraints (physical plausibility bounds) are enforced by mapping to $\boldsymbol{\eta}$-space and rejecting proposals outside the corresponding box $[\boldsymbol{\eta}_{\min},\boldsymbol{\eta}_{\max}]$.

\subsection{Problem Formulation for Parameter Estimation}
\label{subsec:problem}
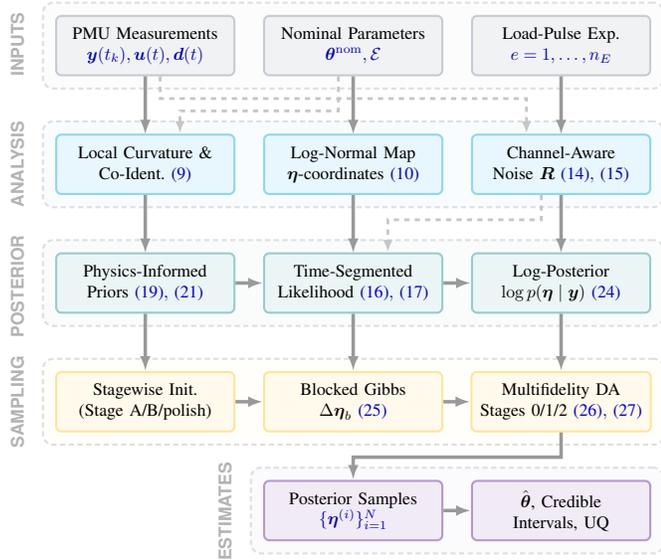
\begin{figure}[t]
\centering
\resizebox{\columnwidth}{!}{%
\begin{tikzpicture}[
>=Latex,
colData/.style={fill=slate!10, draw=slate!40},
colAnal/.style={fill=cyan!8, draw=cyan!40},
colBayes/.style={fill=teal!8, draw=teal!40},
colMCMC/.style={fill=amber!8, draw=amber!40},
colOut/.style={fill=indigo!8, draw=indigo!40},
box/.style={
draw, line width=1.2pt, rounded corners=3pt, align=center,
inner sep=5pt, minimum width=36mm, minimum height=12mm, 
font=\normalsize, fill=white
},
arrow/.style={line width=2.0pt, gray!80, -{Latex[length=3mm,width=2.2mm]}}, 
xarrow/.style={line width=1.6pt, dashed, gray!45, -{Latex[length=2.5mm,width=2mm]}},
grp/.style={draw, rounded corners=6pt, densely dashed, gray!25, line width=1.2pt, inner sep=7pt},
note/.style={font=\bfseries\sffamily, text=gray!60, text depth=0pt},
eq/.style={color=blue!70!black} 
]

\node[box, colData] (pmu) at (0,0) {PMU Measurements \\ \eq{$\boldsymbol{y}(t_k), \boldsymbol{u}(t), \boldsymbol{d}(t)$}};
\node[box, colData] (nom) at (4.2,0) {Nominal Parameters \\ \eq{$\boldsymbol{\theta}^{\mathrm{nom}}, \mathcal{E}$}};
\node[box, colData] (exp) at (8.4,0) {Load-Pulse Exp. \\ \eq{$e=1,\ldots,n_E$}};

\node[box, colAnal] (fisher) at (0,-2.4) {Local Curvature \& \\ Co-Ident. \eq{\eqref{eq:coid}}};
\node[box, colAnal] (reparam) at (4.2,-2.4) {Log-Normal Map \\ $\boldsymbol{\eta}$-coordinates  \eq{\eqref{eq:eta-param}}};
\node[box, colAnal] (noise) at (8.4,-2.4) {Channel-Aware \\ Noise  $\boldsymbol{R}$ \eq{\eqref{eq:sigma_eff}, \eqref{eq:channel_inflate}}};

\node[box, colBayes] (prior) at (0,-4.8) {Physics-Informed \\ Priors  \eq{\eqref{eq:prior_MD}, \eqref{eq:line-lognormal}}};
\node[box, colBayes] (like) at (4.2,-4.8) {Time-Segmented \\ Likelihood  \eq{\eqref{eq:seg_like}, \eqref{eq:multi_exp_like}}};
\node[box, colBayes] (post) at (8.4,-4.8) {Log-Posterior \\ $\log p(\boldsymbol{\eta} \mid \boldsymbol{y})$ \eq{\eqref{eq:logpost_eta}}};

\node[box, colMCMC] (init) at (0,-7.2) {Stagewise Init. \\ {(Stage A/B/polish)}};
\node[box, colMCMC] (prop) at (4.2,-7.2) {Blocked Gibbs \\ $\Delta\boldsymbol{\eta}_b$ \eq{\eqref{eq:block_prop}}};
\node[box, colMCMC] (da) at (8.4,-7.2) {Multifidelity DA \\ Stages 0/1/2  \eq{\eqref{eq:alpha1}, \eqref{eq:alpha2}}};

\node[box, colOut] (samples) at (4.2,-9.4) {Posterior Samples \\ \eq{$\{\boldsymbol{\eta}^{(i)}\}_{i=1}^{N}$}};
\node[box, colOut] (est) at (8.4,-9.4) {$\hat{\boldsymbol{\theta}}$, Credible \\ Intervals, UQ};

\foreach \n in {pmu, nom, exp} \draw[arrow] (\n) -- ++(0,-1.8); 
\foreach \n in {fisher, reparam, noise} \draw[arrow] (\n) -- ++(0,-1.8);

\draw[arrow] (prior) -- (like);
\draw[arrow] (like) -- (post);
\draw[arrow] (init) -- (prop);
\draw[arrow] (prop) -- (da);
\draw[arrow] (samples) -- (est);

\draw[arrow] (post) -- (da);
\draw[arrow] (prior) -- (init);
\draw[arrow] (like) -- (prop);
\draw[arrow] (da.south) -- ++(0,-0.5) -| (samples.north);

\draw[xarrow] ([xshift=-3mm]nom.south) -- ++(0,-0.7) -| ([xshift=7mm]fisher.north);
\draw[xarrow] ([xshift=3mm]pmu.south) -- ++(0,-0.45) -| ([xshift=-7mm]noise.north);
\draw[xarrow] ([xshift=-4mm]noise.south) -- ++(0,-0.5) -| ([xshift=7mm]like.north);

\begin{scope}[on background layer]
\node[grp, fill=slate!2, fit=(pmu)(exp)] (g1) {};
\node[note, rotate=90, xshift=0.8cm, left=0.5cm of g1] {INPUTS};

\node[grp, fill=cyan!2, fit=(fisher)(noise)] (g2) {};
\node[note, rotate=90, xshift=1cm, left=0.5cm of g2] {ANALYSIS};

\node[grp, fill=teal!2, fit=(prior)(post)] (g3) {};
\node[note, rotate=90, xshift=1.1cm, left=0.5cm of g3] {POSTERIOR};

\node[grp, fill=amber!2, fit=(init)(da)] (g4) {};
\node[note, rotate=90, xshift=1.05cm, left=0.5cm of g4] {SAMPLING};

\node[grp, fill=indigo!2, fit=(samples)(est)] (g5) {};
\node[note, rotate=90, xshift=1.1cm, left=0.5cm of g5] {ESTIMATES};
\end{scope}

\end{tikzpicture}
}
\caption{Overview of the DAE-aware Bayesian framework. PMU measurements, nominal parameters, and load-pulse experiments feed identifiability analysis, posterior construction, and multifidelity delayed-acceptance blocked MCMC (Algorithm~1), yielding uncertainty-aware parameter estimates.} \label{fig:framework}
\end{figure}

With the identifiability analysis complete, we formally state the parameter estimation problem:
Given PMU measurements $\{\boldsymbol{y}(t_k)\}_{k=1}^{N_t}$ over time window $[0,T]$, known disturbance/experiment profiles used by the forward model, measurement noise covariance $\boldsymbol{R}\in\mathbb{R}^{p\times p}$ defined by the effective channel-noise model, network topology $\mathcal{E}$ and nominal parameter values $\boldsymbol{\theta}_{\mathrm{nom}}$, the DAE model~\eqref{eq:dae-compact} and measurement model~\eqref{eq:meas}, and the standardized coordinate mapping~\eqref{eq:eta-param}. Find the posterior distribution $p(\boldsymbol{\eta}\mid\boldsymbol{y}_{1:N_t})$ over standardized coordinates $\boldsymbol{\eta}\in\mathbb{R}^{n_\theta}$, point estimates $\hat{\boldsymbol{\eta}}$ (posterior mean) and uncertainty (posterior covariance), full-space parameters $\hat{\boldsymbol{\theta}}=\exp(\boldsymbol{\mu}_\lambda+\boldsymbol{\sigma}_\lambda\odot\hat{\boldsymbol{\eta}})$ with credible intervals. All such that: predicted measurements from the measurement model~\eqref{eq:meas} match observed $\boldsymbol{y}(t_k)$ within noise, parameters satisfy physical constraints (positivity, power-flow feasibility, topology), and uncertainty estimates are summarized by posterior samples and credible intervals. Figure~\ref{fig:framework} summarizes the overall framework.
The next section develops a Bayesian inference framework to solve this problem.

\section{DAE-Aware Bayesian Inference Framework}
\label{sec:bayes}
We now formulate parameter estimation as a Bayesian inference problem. Bayesian methods treat unknown parameters $\boldsymbol{\theta}$ as random variables with a prior distribution $p(\boldsymbol{\theta})$ encoding physical constraints. Given the measurements, Bayes' theorem updates the prior to a posterior distribution:
\begin{align}
p(\boldsymbol{\theta}\mid\boldsymbol{y}_{1:N_t}) = \frac{p(\boldsymbol{y}_{1:N_t}\mid\boldsymbol{\theta})\,p(\boldsymbol{\theta})}{p(\boldsymbol{y}_{1:N_t})} \propto p(\boldsymbol{y}_{1:N_t}\mid\boldsymbol{\theta})\,p(\boldsymbol{\theta}),
\label{eq:bayes_rule}
\end{align}
where the likelihood $p(\boldsymbol{y}_{1:N_t}\mid\boldsymbol{\theta})$ quantifies how well parameters explain the data, and the normalizing constant $p(\boldsymbol{y}_{1:N_t})$ is intractable but cancels in sampling algorithms. The posterior balances data fit (likelihood) against physical plausibility (prior), yielding both point estimates (posterior mean) and calibrated uncertainty (posterior covariance and credible intervals). 
This section develops: \textit{(i)} a channel-aware likelihood function with model-discrepancy noise inflation and time segmentation based on DAE forward solves (Section~\ref{subsec:likelihood}), and \textit{(ii)} physics-informed priors on generator and network parameters (Section~\ref{subsec:priors}). All inference operates in the standardized coordinates $\boldsymbol{\eta}$ from Section~\ref{subsec:identifiability}.

\subsection{Likelihood Construction from PMU Measurements}
\label{subsec:likelihood}

This subsection constructs the likelihood $p(\boldsymbol{y}_{1:N_t}\mid\boldsymbol{\theta})$, quantifying the probability of observing PMU measurements $\boldsymbol{y}_{1:N_t}$ given parameters $\boldsymbol{\theta}$. We assume the implementation has access to known transient-experiment data recorded during one or more transient events over time window $[0,T]$: PMU measurements including selected rectangular voltage components together with generator frequency deviations $\boldsymbol{y}(t_k)$ at discrete times $\{t_k\}_{k=1}^{N_t}$ (typically 30--120 Hz sampling), and known disturbance/load-pulse profiles $\boldsymbol{d}(t)$ used to excite the system in each experiment.

The measurement model~\eqref{eq:meas} relates observed $\boldsymbol{y}(t_k)$ to the predicted measurement channels generated from the forward solution. Given these measurements, we define the residual vector at time $t_k$:
\begin{align}
\boldsymbol{r}_k(\boldsymbol{\theta}) := \boldsymbol{y}(t_k) - \boldsymbol{h}_{\mathrm{meas}}\!\left(\boldsymbol{x}_d(t_k;\boldsymbol{\theta}),\boldsymbol{x}_a(t_k;\boldsymbol{\theta})\right) \in\mathbb{R}^{p}.
\label{eq:residual}
\end{align}
Under the Gaussian noise assumption, the likelihood for a single measurement is $p(\boldsymbol{y}(t_k)\mid\boldsymbol{\theta}) = \mathcal{N}\!\left(\boldsymbol{y}(t_k);\,\boldsymbol{h}_{\mathrm{meas}}\!\left(\boldsymbol{x}_d(t_k;\boldsymbol{\theta}),\boldsymbol{x}_a(t_k;\boldsymbol{\theta})\right),\boldsymbol{R}\right)$. Assuming independence across time steps, the joint log-likelihood over all $N_t$ samples is: 
\begin{equation}
\begin{aligned}
\log p(\boldsymbol{y}_{1:N_t}\mid\boldsymbol{\theta})
= -\tfrac{1}{2}\sum_{k=1}^{N_t}\boldsymbol{r}_k(\boldsymbol{\theta})^\top \boldsymbol{R}^{-1}\boldsymbol{r}_k(\boldsymbol{\theta}) \\- \tfrac{N_t p}{2}\log(2\pi) - \tfrac{N_t}{2}\log|\boldsymbol{R}|.
\label{eq:gauss_like}
\end{aligned}
\end{equation}
The quadratic form $\boldsymbol{r}_k^\top\boldsymbol{R}^{-1}\boldsymbol{r}_k$ penalizes residuals weighted by inverse noise covariance so that large mismatches in low-noise channels are penalized, while noisy channels contribute less.
\subsubsection{Channel-specific model-discrepancy inflation}
\label{subsubsec:model_discrepancy}
A critical practical issue is that the forward DAE model is an approximation of reality: numerical integration introduces discretization error, the fourth-order generator model omits subtransient and saturation effects, and the linearized governor/exciter dynamics are simplified. If the effective noise covariance $\boldsymbol{R}$ reflects only measurement noise, the resulting posterior becomes unrealistically sharp, the likelihood dominates the prior by many orders of magnitude, and the MCMC chain cannot explore the posterior effectively.

To address this, we introduce a model-discrepancy term that inflates the effective noise variance per channel to honestly account for model approximation error. Let $\sigma_j^{\mathrm{meas}}$ denote the measurement noise standard deviation for channel $j$ and let $\hat{s}_j$ denote the empirical standard deviation of the clean signal in channel $j$ computed from the corresponding clean simulated channel data. The effective noise standard deviation is
\begin{equation} \small
\begin{aligned} 
\sigma_j^{\mathrm{eff}} = \kappa_j\sqrt{(\sigma_j^{\mathrm{meas}})^2 + (\sigma_j^{\mathrm{model}})^2},\quad \sigma_j^{\mathrm{model}}=\max\!\big(\sigma_{\mathrm{floor}},\;\rho\,\hat{s}_j\big),
\label{eq:sigma_eff}
\end{aligned}
\end{equation}
where $\rho$ is a relative model-discrepancy fraction (e.g., $\rho=0.02$ corresponds to 2\% of the signal standard deviation), $\sigma_{\mathrm{floor}}$ prevents division by zero for near-constant channels, and $\kappa_j$ is a channel-class-specific inflation factor. We set $\kappa_j$ differently for voltage and frequency channels:
\begin{align}
\kappa_j = \begin{cases} \kappa_{\mathrm{volt}} & j\in\mathcal{C}_{\mathrm{volt}}, \\[3pt] \kappa_{\mathrm{freq}} & j\in\mathcal{C}_{\mathrm{freq}},\end{cases}
\label{eq:channel_inflate}
\end{align}
where $\mathcal{C}_{\mathrm{volt}}$ and $\mathcal{C}_{\mathrm{freq}}$ partition the measurement channels into voltage and frequency classes. The rationale is that generator dynamic parameters ($\boldsymbol{M},\boldsymbol{D}$) are strongly identified from frequency channels even under aggressive inflation ($\kappa_{\mathrm{freq}}\gg 1$), while network parameters ($\boldsymbol{r},\boldsymbol{x}$) are primarily identified from voltage channels and require more conservative inflation ($\kappa_{\mathrm{volt}}$ moderate). This channel-specific treatment avoids the failure mode where uniform inflation drowns the voltage signal and severely weakens practical network-parameter recovery, while still preventing the posterior from becoming pathologically sharp.
The effective noise covariance used in~\eqref{eq:gauss_like} is then $\boldsymbol{R}=\mathrm{diag}((\sigma_1^{\mathrm{eff}})^2,\ldots,(\sigma_p^{\mathrm{eff}})^2)$. The key diagnostic for tuning the inflation is the ratio $|\ell_{\mathrm{data}}|/|\ell_{\mathrm{prior}}|$ at the posterior center, which should be in the range $10^3$--$10^6$; values above $10^{10}$ indicate that the prior is irrelevant and the posterior is unrealistically sharp.
\subsubsection{Time-segmented likelihood}
Since electromechanical transients exhibit a useful time-scale separation --- \textit{(i)} early-time frequency swing encodes inertia, \textit{(ii)} mid-time oscillation decay encodes damping, \textit{(iii)} late-time settling and voltage distribution encode network admittances --- we exploit this structure by partitioning the time index set \(\{1,\dots,N_t\}\) into an unweighted baseline segment together with \(\mathcal{T}_M\), \(\mathcal{T}_D\), \(\mathcal{T}_Y\), corresponding to the inertia-dominant, damping-dominant, and network-settling windows respectively. Each non-baseline window receives channel-specific weights $w_{jk}$ that emphasize the informative channels for that time segment. Samples outside the pulse-driven windows retain unit weight. 
\begin{align}
\small
\log p(\boldsymbol{y}_{1:N_t}\mid\boldsymbol{\eta})
&= -\tfrac{1}{2}\sum_{k=1}^{N_t}\sum_{j=1}^{p} w_{jk}\, \frac{r_{jk}^2(\boldsymbol{\theta}(\boldsymbol{\eta}))}{(\sigma_j^{\mathrm{eff}})^2}
+ \mathrm{const},
\label{eq:seg_like}
\end{align}
where $r_{jk}$ is the $j$-th component of the residual $\boldsymbol{r}_k$ and the weights $w_{jk}$ encode the time-channel segmentation. Specifically, for each load-pulse event the time axis is partitioned into an inertia window (immediately after pulse onset, duration 0.35\,s), a damping window (after pulse removal, duration 1.2\,s), and a network-settling window (the remaining time until the next pulse). Within each window, frequency channels receive a mild emphasis during the inertia and damping segments (e.g., $w_{\mathrm{freq},M}=1.3$, $w_{\mathrm{freq},D}=1.2$), while voltage channels receive mild emphasis during the network-settling segment (e.g., $w_{\mathrm{volt},Y}=1.2$). The weights are normalized to preserve the average information scale. This segmentation is complementary to the channel-specific inflation~\eqref{eq:channel_inflate}: the inflation addresses the \emph{channel} dimension of information content, while segmentation addresses the \emph{temporal} dimension.
\subsubsection{Multi-experiment likelihood}
When multiple transient experiments $e=1,\ldots,n_E$ are available (e.g., load pulses applied at different buses), the joint likelihood is the product of per-experiment likelihoods. Each experiment provides an independent set of residuals $\boldsymbol{r}_k^{(e)}(\boldsymbol{\theta})$, and total log-likelihood sums over experiments and time steps:
\begin{align} \small
\log p(\boldsymbol{y}\mid\boldsymbol{\theta}) = -\tfrac{1}{2}\sum_{e=1}^{n_E}\sum_{k=1}^{N_t}\sum_{j=1}^{p} w_{jk}^{(e)}\, \frac{\big(r_{jk}^{(e)}(\boldsymbol{\theta})\big)^2}{(\sigma_j^{\mathrm{eff}})^2} + \mathrm{const}.\label{eq:multi_exp_like}
\end{align}
Multiple experiments with diverse excitation patterns improve identifiability by exciting different modes of the system. In our implementation, load-pulse trains are applied sequentially at different buses so that each experiment emphasizes different generators and network paths.

To evaluate the likelihood for any candidate $\boldsymbol{\theta}$, we must compute the predicted measurements through the measurement model~\eqref{eq:meas} by solving the forward DAE problem. This requires two steps for each $\boldsymbol{\theta}$: \textit{(i)} solve the nonlinear system~\eqref{eq:pfcons} to find initial conditions $(\boldsymbol{x}_d^\star,\boldsymbol{x}_a^\star)$ satisfying $\boldsymbol{f}(\boldsymbol{x}^\star,\boldsymbol{u}_0;\boldsymbol{\theta})=\boldsymbol{0}$ and $\boldsymbol{g}(\boldsymbol{x}^\star,\boldsymbol{d}_0;\boldsymbol{\theta})=\boldsymbol{0}$ for pre-event setpoints $(\boldsymbol{u}_0,\boldsymbol{d}_0)$ and \textit{(ii)} integrate the index-1 DAE~\eqref{eq:dae-compact} from $t=0$ to $T$ using the disturbance inputs $\boldsymbol{u}(t)$ and $\boldsymbol{d}(t)$, starting from $(\boldsymbol{x}_d^\star,\boldsymbol{x}_a^\star)$. This yields trajectories $(\boldsymbol{x}_d(t;\boldsymbol{\theta}),\boldsymbol{x}_a(t;\boldsymbol{\theta}))$, from which we construct predicted measurement channels.

These forward solves are repeated for every parameter candidate evaluated during MCMC sampling (Section~\ref{sec:pf_mcmc}). To form the likelihood in the standardized coordinates $\boldsymbol{\eta}$ from~\eqref{eq:eta-param}, we reconstruct physical parameters via $\boldsymbol{\theta}(\boldsymbol{\eta})=\exp(\boldsymbol{\mu}_\lambda+\boldsymbol{\sigma}_\lambda\odot\boldsymbol{\eta})$, perform the forward solve at $\boldsymbol{\theta}(\boldsymbol{\eta})$, and evaluate
\begin{align}
\log p(\boldsymbol{y}_{1:N_t}\mid\boldsymbol{\eta}) = \log p(\boldsymbol{y}_{1:N_t}\mid\boldsymbol{\theta}(\boldsymbol{\eta})).
\label{eq:like_alpha}
\end{align}
All subsequent computations use $\boldsymbol{\eta}$ as the inference variable, with $\boldsymbol{\theta}(\boldsymbol{\eta})$ reconstructed only when calling the DAE solver.

Having constructed channel-aware, time-segmented likelihood, we next introduce the physics-informed priors that, combined with this likelihood through Bayes' rule~\eqref{eq:bayes_rule}, allow us to obtain posterior distributions for parameter inference.

\subsection{Physics-Informed Priors}
\label{subsec:priors}
This section defines the weakly-informative priors $p(\boldsymbol{\theta})$ that avoid biasing estimates toward specific numerical values while enforcing hard physical constraints.

\subsubsection{Generator parameter priors}

Inertia constants $M_i$ and damping coefficients $D_i$ are strictly positive and typically uncertain by multiplicative factors (due to factors such as generator aging or rewinding). We model this multiplicative uncertainty via log-normal distributions. For each generator $i=1,\dots,n_G$, let $M_i^{\text{nom}}$ and $D_i^{\text{nom}}$ be nominal values from manufacturer data or system records. The log-normal prior is:
\begin{subequations}\label{eq:prior_MD}
\begin{align}
p(M_i) &= \mathcal{LN}(M_i;\,\mu_{M_i},\sigma_M^2)\,\mathbb{I}_{[M_{\min},M_{\max}]}(M_i), \\
p(D_i) &= \mathcal{LN}(D_i;\,\mu_{D_i},\sigma_D^2)\,\mathbb{I}_{[D_{\min},D_{\max}]}(D_i),
\end{align}
\end{subequations}
where $\mathcal{LN}(x;\mu,\sigma^2)$ denotes the log-normal density with $\log(x)\sim\mathcal{N}(\mu,\sigma^2)$, location parameter $\mu_{M_i}=\log(M_i^{\text{nom}})$, and scale $\sigma_M$ (similarly for damping). The indicator function $\mathbb{I}_{[M_{\min},M_{\max}]}(M_i)$ truncates to physically plausible bounds. Assuming independence across generators:
\begin{align}
p(\boldsymbol{M},\boldsymbol{D}) = \prod_{i=1}^{n_G} p(M_i)\,p(D_i).
\end{align}
The dispersion parameters $\sigma_M$ and $\sigma_D$ are set so that the 95\% prior spread corresponds approximately to a prescribed multiplicative range around the nominal values. Exact prior widths are specified in the case studies. This prior is weakly-informative, it allows substantial deviation from nominals while preventing unphysical values such as negative inertia.

\subsubsection{Network parameter priors}

Rather than estimating entries of the bus admittance matrix directly, we estimate the physical series branch parameters---resistances $r_\ell$ and reactances $x_\ell$---for each line $\ell\in\mathcal{E}$ with nonzero values. The bus admittance matrix is then assembled from these branch parameters via~\eqref{eq:Y-assembly}. This approach exploits the known network topology to enforce symmetry, sign patterns, and Kirchhoff's current law automatically, and reduces the number of free network parameters from $|\mathrm{nnz}(\boldsymbol{G})|+|\mathrm{nnz}(\boldsymbol{B})|$ to $n_r+n_x$ where $n_r$ and $n_x$ count the nonzero-resistance and nonzero-reactance branches respectively.

We model multiplicative uncertainty in branch parameters via log-normal perturbations. For each branch $\ell\in\mathcal{E}$ with nonzero resistance or reactance:
\begin{equation}\label{eq:line-lognormal}
\begin{aligned}
&r_\ell = r_\ell^{\mathrm{nom}}\exp(\sigma_r\,\eta_{\ell,r}),\\
&x_\ell = x_\ell^{\mathrm{nom}}\exp(\sigma_x\,\eta_{\ell,x}),\quad
&\eta_{\ell,r},\eta_{\ell,x}\sim \mathcal{N}(0,1).
\end{aligned}
\end{equation}
Here, $\eta_{\ell,r}$ and $\eta_{\ell,x}$ are the standardized coordinates from~\eqref{eq:eta-param} restricted to network parameters. This preserves strict positivity of $r_\ell$ and $x_\ell$ by construction. Assuming independence across branches, the joint prior is:
\begin{align}
p(\boldsymbol{r},\boldsymbol{x}) = \prod_{\ell:\,r_\ell>0} p(r_\ell)\,\prod_{\ell:\,x_\ell>0}p(x_\ell).
\end{align}
Combining~\eqref{eq:line-lognormal} with~\eqref{eq:Y-assembly} guarantees that all sampled branch parameters yield valid $\boldsymbol{Y}_{\mathrm{bus}}$ matrices with correct topology, symmetry, and sign patterns.

\subsubsection{Power-flow feasibility constraint}

Beyond priors, we enforce a practical power-flow feasibility constraint: parameter proposals for which the Newton-Raphson solve of~\eqref{eq:pfcons} fails to converge are immediately rejected. In implementation, successful convergence of this initialization solve is used as the feasibility test so such proposals are assigned zero posterior support and discarded without evaluating likelihood. Since each MCMC iteration already begins with a power-flow solve as part of the DAE initialization (Section~\ref{sec:pf_mcmc}), this check adds no extra computational cost.

\subsubsection{Combined posterior in standardized coordinates}

The full prior combines generator, network, and feasibility terms:
$p(\boldsymbol{\theta}) = p(\boldsymbol{M},\boldsymbol{D})\,p(\boldsymbol{r},\boldsymbol{x})\,p_{\mathrm{PF}}(\boldsymbol{\theta}).$
Using the reparameterization $\boldsymbol{\theta}(\boldsymbol{\eta})=\exp(\boldsymbol{\mu}_\lambda+\boldsymbol{\sigma}_\lambda\odot\boldsymbol{\eta})$ from~\eqref{eq:eta-param}, the posterior in standardized coordinates is:
\begin{align}
p(\boldsymbol{\eta}\mid\boldsymbol{y}_{1:N_t})
\propto p(\boldsymbol{y}_{1:N_t}\mid\boldsymbol{\eta})\,
p_{\mathrm{PF}}(\boldsymbol{\theta}(\boldsymbol{\eta}))\,
p(\boldsymbol{\eta}).
\label{eq:posterior_alpha}
\end{align}

In $\boldsymbol{\eta}$-coordinates, the Gaussian part of the log-prior contribution from~\eqref{eq:prior_MD} and~\eqref{eq:line-lognormal} reduces to $-\tfrac{1}{2}\|\boldsymbol{\eta}\|^2$, so that within the admissible box the log-posterior is
\begin{align}
\log p(\boldsymbol{\eta}\mid\boldsymbol{y}) = \underbrace{-\tfrac{1}{2}\sum_{e,k,j} w_{jk}^{(e)}\frac{(r_{jk}^{(e)})^2}{(\sigma_j^{\mathrm{eff}})^2}}_{\text{log-likelihood}} \underbrace{-\,\tfrac{1}{2}\|\boldsymbol{\eta}\|^2}_{\text{log-prior}} + \mathrm{const},
\label{eq:logpost_eta}
\end{align}
subject to $\boldsymbol{\eta}\in[\boldsymbol{\eta}_{\min},\boldsymbol{\eta}_{\max}]$ and power-flow feasibility $p_{\mathrm{PF}}(\boldsymbol{\theta}(\boldsymbol{\eta}))=1$.
Having constructed the channel-aware, time-segmented likelihood~\eqref{eq:multi_exp_like} and physics-informed priors (Section~\ref{subsec:priors}), and with the posterior distribution~\eqref{eq:posterior_alpha} being specified. The following section describes an MCMC sampling algorithm to explore this posterior efficiently while respecting DAE constraints.

\section{Multifidelity Delayed-Acceptance MCMC}
\label{sec:pf_mcmc}

Sampling from the posterior~\eqref{eq:posterior_alpha} requires an algorithm that generates parameter samples $\{\boldsymbol{\eta}^{(i)}\}_{i=1}^N$ distributed according to $p(\boldsymbol{\eta}\mid\boldsymbol{y}_{1:N_t})$. Markov Chain Monte Carlo (MCMC) achieves this by constructing a Markov chain whose stationary distribution equals the target posterior. The Metropolis-Hastings algorithm~\cite{metropolis1953equation,hastings1970monte} is a foundational MCMC method that iterates: \textit{(i)} Propose a new state $\boldsymbol{\eta}'$ from current state $\boldsymbol{\eta}$ using a proposal distribution $q(\boldsymbol{\eta}'\mid\boldsymbol{\eta})$, \textit{(ii)} Accept the proposal with probability $\alpha_{\text{MH}} = \min\{1,\, \frac{p(\boldsymbol{\eta}')q(\boldsymbol{\eta}\mid\boldsymbol{\eta}')}{p(\boldsymbol{\eta})q(\boldsymbol{\eta}'\mid\boldsymbol{\eta})}\}$, \textit{(iii)} If accepted, set $\boldsymbol{\eta}\leftarrow\boldsymbol{\eta}'$, otherwise retain $\boldsymbol{\eta}$.
For symmetric proposals, this simplifies to accepting with probability $\min\{1,\, p(\boldsymbol{\eta}')/p(\boldsymbol{\eta})\}$. Over many iterations, the chain's empirical distribution converges to the posterior regardless of initialization enabling estimation of posterior statistics (mean, covariance, quantiles) from samples.

Standard Metropolis proposals $\boldsymbol{\eta}'\sim\mathcal{N}(\boldsymbol{\eta},\boldsymbol{\Sigma})$ in high-dimensional parameter spaces face two challenges for DAE systems: \textit{(i)} proposals often violate power-flow feasibility (the equilibrium system~\eqref{eq:pfcons} has no solution), causing DAE initialization failures and zero acceptance, and \textit{(ii)} each likelihood evaluation requires expensive DAE integration, making low acceptance rates computationally prohibitive. This section develops two key enhancements that address these issues while preserving the correct stationary distribution: blocked Gibbs proposals with adaptive covariance learning (Section~\ref{subsec:blocked}), and multifidelity delayed acceptance using a coarse-grid DAE surrogate (Section~\ref{subsec:delayed}).
\subsection{Blocked Gibbs Proposals with Adaptive Covariance}
\label{subsec:blocked}
Rather than proposing all $n_\theta$ parameters simultaneously, we partition the parameter vector into physically meaningful blocks and cycle through them. This blocked Gibbs strategy dramatically improves acceptance rates by reducing the effective dimensionality of each proposal. The blocks are: \textit{(i)} Dynamics block $\mathcal{B}_{\mathrm{dyn}}$: inertias and dampings $(\boldsymbol{M},\boldsymbol{D})$ --- $2n_G$ parameters, \textit{(ii)} Resistance block $\mathcal{B}_r$: nonzero branch resistances $\boldsymbol{r}$ --- $n_r$ parameters, and \textit{(iii)} Reactance block $\mathcal{B}_x$: nonzero branch reactances $\boldsymbol{x}$ --- $n_x$ parameters.

At each MCMC iteration, one block $b\in\{\mathcal{B}_{\mathrm{dyn}},\mathcal{B}_r,\mathcal{B}_x\}$ is selected cyclically. A Gaussian proposal is drawn for selected block's coordinates $\boldsymbol{\eta}_b$, with other coordinates held fixed:
\begin{align}
\Delta\boldsymbol{\eta}_b \sim \mathcal{N}\!\big(\boldsymbol{0},\;s_b^2\cdot\tfrac{2.38^2}{d_b}\cdot\boldsymbol{L}_b\boldsymbol{L}_b^\top\big),
\label{eq:block_prop}
\end{align}
where $d_b=|\mathcal{B}_b|$ is the block dimension, $\boldsymbol{L}_b$ is the Cholesky factor of the block's proposal covariance, and $s_b$ is an adaptive scale factor. The factor $2.38^2/d_b$ is the asymptotically optimal Metropolis scaling. The proposal covariance is initialized from a local finite-difference Hessian approximation at the posterior center (obtained from a short stagewise optimization), then adapted during burn-in using empirical chain covariance blended with the initial estimate for stability.

During the burn-in phase, the scale $s_b$ is adapted every $n_{\mathrm{adapt}}$ iterations using the Robbins--Monro rule: $\log s_b \leftarrow \log s_b + \gamma_t(\hat{a}_b - a_{\mathrm{target}})$, where $\hat{a}_b$ is the recent acceptance rate for block $b$, $a_{\mathrm{target}}\approx 0.24$ is the target acceptance, and $\gamma_t$ is a decaying step size. Simultaneously, the proposal covariance $\boldsymbol{L}_b\boldsymbol{L}_b^\top$ is updated as a convex combination of the empirical covariance from recent chain history and the initial Hessian-based covariance, preventing both over-adaptation to early transients and forgetting of local curvature information. Adaptation is frozen after burn-in to preserve detailed balance.

\subsection{Multifidelity Delayed Acceptance with Coarse-Grid DAE}
\label{subsec:delayed}

Even with blocked proposals, each accepted proposal requires full DAE integration over $[0,T]$ to evaluate the likelihood~\eqref{eq:gauss_like}. At typical acceptance rates of 20\%, this means a large fraction of expensive DAE solves are wasted on rejected proposals. Delayed acceptance~\cite{christen2005markov} addresses this by introducing a three-stage filter: Stage~0 screens for power-flow feasibility by running Newton--Raphson at the proposed parameters, rejecting infeasible proposals at negligible cost. Stage~1 evaluates a cheap coarse-fidelity likelihood by integrating the DAE on a decimated time grid with relaxed solver tolerances. Stage~2 computes the exact likelihood via the full-fidelity DAE integration, but only for proposals that pass both previous stages. A Metropolis correction factor ensures the chain targets the exact posterior despite the approximate screening.

The acceptance probabilities are then given by:
\begin{enumerate}[leftmargin=*,itemsep=0pt]
\item \textit{Stage 0 (power-flow screen):} Given proposed $\boldsymbol{\eta}'$, reconstruct $\boldsymbol{\theta}'=\boldsymbol{\theta}(\boldsymbol{\eta}')$ and solve the power flow~\eqref{eq:pfcons}. If Newton--Raphson diverges, reject immediately.
\item \textit{Stage 1 (coarse DAE):} Integrate the DAE on a decimated time grid with relaxed solver tolerances to obtain a cheap approximate log-posterior $\widetilde{\ell}(\boldsymbol{\eta}')$. Accept with probability:
\begin{align}
\alpha_1 = \min\!\left\{1,\;
\exp\!\big(\widetilde{\ell}(\boldsymbol{\eta}')-\widetilde{\ell}(\boldsymbol{\eta})\big)
\right\}.
\label{eq:alpha1}
\end{align}
If rejected, skip to next iteration without the expensive exact DAE solver.
\item \textit{Stage 2 (exact):} If Stage 1 accepted, compute the exact log-posterior $\ell(\boldsymbol{\eta}')$ (and compare it with the current exact and coarse values) via full-fidelity DAE integration. Accept with correction probability:
\begin{align} \small
\alpha_2 = \min\!\left\{1,\;
\exp\!\Big[(\ell(\boldsymbol{\eta}')-\ell(\boldsymbol{\eta}))-(\widetilde{\ell}(\boldsymbol{\eta}')-\widetilde{\ell}(\boldsymbol{\eta}))\Big]
\right\}.
\label{eq:alpha2}
\end{align}
The correction factor accounts for the discrepancy between the exact and coarse log-posterior increments, ensuring unbiased sampling.
\end{enumerate}
The composition of Stages~1 and~2 defines a valid delayed-acceptance Metropolis kernel with the exact posterior as invariant distribution, while the number of proposals reaching Stage~2 (the expensive exact DAE solve) is substantially reduced. The key requirement for efficient delayed acceptance is that the coarse and exact likelihoods are well correlated: if the coarse model reliably distinguishes good from bad proposals, Stage~1 acceptance rates are high and few exact solves are wasted. In our implementation, the coarse model uses the same physical DAE but with a sparser time grid and slightly relaxed ODE solver tolerances, ensuring high correlation with the exact likelihood while being several times cheaper to evaluate.

\subsection{Stagewise Initialization}
\label{subsec:init}

Rather than starting MCMC from the nominal parameter values (which may be far from the posterior mode), we use a fast stagewise nonlinear least-squares initialization to place the chain in a high-probability region. In Stage~A, generator parameters ($\boldsymbol{M},\boldsymbol{D}$) are optimized against frequency-channel residuals with network parameters fixed at nominal values. In Stage~B, network parameters ($\boldsymbol{r},\boldsymbol{x}$) are optimized against all-channel residuals with dynamics fixed at Stage~A values. An optional joint polish refines all parameters simultaneously. This initialization provides: \textit{(i)} a good starting point for MCMC that dramatically reduces burn-in, and \textit{(ii)} a local Gauss-Newton curvature approximation formed from a finite-difference Jacobian of the stacked residual vector at the selected stagewise center, used to initialize the block proposal covariances in~\eqref{eq:block_prop}.

\subsection{Complete Algorithm}
\label{subsec:algorithm}
Algorithm 1 summarizes all steps: stagewise initialization, blocked proposals with adaptive covariance, and multifidelity delayed acceptance.
\begin{algorithm}
\DontPrintSemicolon
\small
\SetAlFnt{\small}
\SetAlCapFnt{\small}
\SetAlCapNameFnt{\small}
\SetInd{0.4em}{0.8em} 
\caption{\blue{Multifidelity Delayed-Acceptance MCMC}}
\SetKwInOut{Input}{Inputs}
\SetKwInOut{Output}{Outputs}
\Input{Data $(\boldsymbol{y}_{1:N_t},\blue{\boldsymbol{d}(t)})$, \blue{nominal $\boldsymbol{\theta}^{\mathrm{nom}}$, noise model $\boldsymbol{R}$}}
\Output{Posterior samples $\{\blue{\boldsymbol{\eta}}^{(i)}\}$, reconstruct $\boldsymbol{\theta}^{(i)}=\blue{\exp(\boldsymbol{\mu}_\lambda+\boldsymbol{\sigma}_\lambda\odot\boldsymbol{\eta}^{(i)})}$}
\BlankLine
\blue{\textbf{Initialize:} Run Stage A/B/joint optimization $\to \boldsymbol{\eta}^{(0)}$, compute local \blue{Gauss--Newton curvature} $\to$ block covariances $\{\boldsymbol{\Sigma}_b\}$}\;
\BlankLine
\For{$i=1$ \KwTo $N_{\text{burn}} + N_{\text{samp}}\,N_{\text{thin}}$}{
Select block $b \in \{\blue{\mathcal{B}_{\mathrm{dyn}}, \mathcal{B}_r, \mathcal{B}_x}\}$ \blue{(blockwise schedule)}\;
Draw $\Delta\blue{\boldsymbol{\eta}}_b \sim \mathcal{N}(\boldsymbol{0}, \blue{s_b^2 \cdot \tfrac{2.38^2}{d_b} \cdot} \boldsymbol{\Sigma}_b)$, set $\blue{\boldsymbol{\eta}}' = \blue{\boldsymbol{\eta}} + \Delta\blue{\boldsymbol{\eta}}$\;
\blue{\textbf{Box check:}} \If{$\boldsymbol{\eta}' \notin [\boldsymbol{\eta}_{\min}, \boldsymbol{\eta}_{\max}]$}{
Reject and \textbf{continue}\;}
\BlankLine
\textbf{\blue{Stage 0}:} \blue{Solve power flow at $\boldsymbol{\theta}(\boldsymbol{\eta}')$}; if failed, reject and \textbf{continue}\;
\textbf{Stage 1:} Evaluate \blue{coarse DAE} $\widetilde{\ell}(\blue{\boldsymbol{\eta}}')$, compute $\alpha_1$ via~\eqref{eq:alpha1}\;
\If{$\text{rand}() < \alpha_1$}{
\textbf{Stage 2:} \blue{Integrate full-fidelity DAE}, compute $\ell(\blue{\boldsymbol{\eta}}')$\;
Compute $\alpha_2$ via~\eqref{eq:alpha2}\;
\If{$\text{rand}() < \alpha_2$}{
$\blue{\boldsymbol{\eta}} \leftarrow \blue{\boldsymbol{\eta}}'$, $\ell \leftarrow \ell'$ \tcc*[r]{Accept proposal}}}
\BlankLine
\If{$i > N_{\text{burn}}$ \textbf{and} $\bmod(i-N_{\text{burn}}, N_{\text{thin}})=0$}{
Store the current state as the next posterior sample $\blue{\boldsymbol{\eta}}^{(m)} = \blue{\boldsymbol{\eta}}$\;}
\If{$\bmod(i, \blue{n_{\mathrm{adapt}}}) = 0$ \textbf{and} $i \le N_{\text{burn}}$}{
Adapt $\blue{s_b}$ and $\boldsymbol{\Sigma}_b$ to target \blue{$\approx$24\%} \tcc*[r]{Tuning}}}
\end{algorithm}
This MCMC framework delivers: \textit{(i)} power-flow feasibility enforced at Stage~0 of every iteration at negligible cost; \textit{(ii)} computational efficiency through multifidelity delayed acceptance that filters proposals via cheap coarse-grid DAE solves before committing to expensive exact evaluations; \textit{(iii)} posterior-targeting sampling after burn-in since adaptation is frozen and the resulting delayed-acceptance Metropolis kernel satisfies detailed balance; and \textit{(iv)} efficient exploration via blocked Gibbs proposals in standardized log-coordinates with adaptive covariance learning. 
The following section presents case studies and the exact simulation setup.

\section{Case Studies}
\label{sec:case_studies}

This section validates the framework on the IEEE 9-bus and 39-bus systems, first describing the test setup, then presenting co-identifiability diagnostics, and finally comparing joint versus decoupled estimation to demonstrate the practical impact of the DAE coupling quantified in Section~\ref{subsec:coident}.

\subsection{System Description and Experimental Design}

We consider the IEEE 9-bus system with $n_G=3$ synchronous generators (buses 1, 2, 3), three load buses (5, 7, 9), and nine transmission lines. Each generator uses the fourth-order model from Section~\ref{sec:prelim} with turbine-governor time constant $T_{\mathrm{ch}}=0.2$\,s and droop $R_d=0.02\cdot 2\pi$, under a linear stabilizing controller $\boldsymbol{K}$.
Among the nine branches, three generator-to-network transformer branches have zero series resistance, while the remaining six have nonzero resistance; all nine have nonzero reactance. Thus, the unknown vector~\eqref{eq:theta} contains $3$ inertias, $3$ dampings, $6$ branch resistances, and $9$ branch reactances, a total of $n_\theta=21$ parameters. Bus admittance matrix $\boldsymbol{Y}_{\mathrm{bus}}$ is assembled from these branch parameters via~\eqref{eq:Y-assembly}, with all other branch data fixed at nominal values.

The measurement vector has dimension $p=17$ and consists of $14$ rectangular voltage components $(V_r,V_i)$ at monitored buses and $3$ generator frequency-deviation channels, as defined by~\eqref{eq:meas}. The system is excited by four load-pulse experiments at buses 5, 7, and 9: three single-bus pulse trains and one paired-bus alternating pattern. Pulse durations are 0.4--0.45\,s and amplitudes are 8--22\% of local load; reactive perturbations are scaled to preserve the local $Q/P$ ratio. The simulation horizon is $T=10$\,s with $\Delta t=0.01$\,s, and after $1\!:\!16$ decimation the fit grid contains $N_t=63$ time points per experiment.
Synthetic PMU measurements are generated by simulating the DAE~\eqref{eq:dae-compact} at the true parameters, obtained by perturbing nominal values by up to $\pm 6\%$ for $\boldsymbol{M}$, $\pm 10\%$ for $\boldsymbol{D}$, and $\pm 8\%$ for $\boldsymbol{r}$ and $\boldsymbol{x}$, and adding Gaussian noise at 25\,dB SNR. The true generator parameters are listed in Tab.~\ref{tab:true_params}.

\begin{table}[h]
\centering
\caption{Nominal and true generator parameters (IEEE 9-bus)}\label{tab:true_params}
\resizebox{\columnwidth}{!}{%
\begin{tabular}{lcccc}
\hline
Parameter & Nominal & Gen 1 & Gen 2 & Gen 3 \\
\hline
$M_i$ [s] & 0.236/0.064/0.030 & 0.231 & 0.066 & 0.031 \\
$D_i$ [pu] & 1.92/0.50/0.20 & 1.958 & 0.472 & 0.194 \\
\hline
\end{tabular}}
\end{table}

\subsection{Tuning and Configuration}
The framework has a number of tunable quantities, as is common in Bayesian MCMC methods. We highlight the most consequential choices. \textit{Priors:} log-normal priors~\eqref{eq:prior_MD}--\eqref{eq:line-lognormal} are centered at nominal values with 95\% widths of $\pm 30\%$ for $\boldsymbol{M}$, $\pm 60\%$ for $\boldsymbol{D}$, and $\pm 25\%$ for $\boldsymbol{r},\boldsymbol{x}$; box constraints truncate proposals at $\pm 50\%$, $\pm 90\%$, and $\pm 45\%$, respectively. \textit{Noise model:} the channel-aware inflation~\eqref{eq:sigma_eff}--\eqref{eq:channel_inflate} uses $\rho=0.02$, $\kappa_{\mathrm{freq}}=15$, and $\kappa_{\mathrm{volt}}=5$, yielding $|\ell_{\mathrm{data}}|/|\ell_{\mathrm{prior}}|\approx 10^5$ at the posterior center. \textit{Time segmentation:} each pulse response is split into inertia (0.35\,s after pulse onset), damping (1.2\,s after pulse removal), and network-settling windows, with mild weights $w_{\mathrm{freq},M}=1.3$, $w_{\mathrm{freq},D}=1.2$, and $w_{\mathrm{volt},Y}=1.2$, normalized to preserve the average information scale. \textit{MCMC:} the blocked Gibbs sampler cycles through dynamics, resistance, and reactance blocks; burn-in uses 3000 iterations with adaptation every 50 iterations targeting 24\% acceptance, followed by 2000 production samples with thinning factor 2. The delayed-acceptance coarse model uses a decimation factor of 24 (vs.\ 16 for the exact grid) with relaxed solver tolerances, and Stage~0 enforces power-flow feasibility. \textit{Initialization:} a stagewise procedure first fits $(\boldsymbol{M},\boldsymbol{D})$ to frequency residuals, then $(\boldsymbol{r},\boldsymbol{x})$ to all channels, followed by a short joint polish; the resulting point initializes the chain and the local Gauss--Newton curvature used in~\eqref{eq:block_prop}.

\subsection{Co-Identifiability Analysis}

\begin{figure}[t]
\centering
\includegraphics[width=0.8\linewidth]{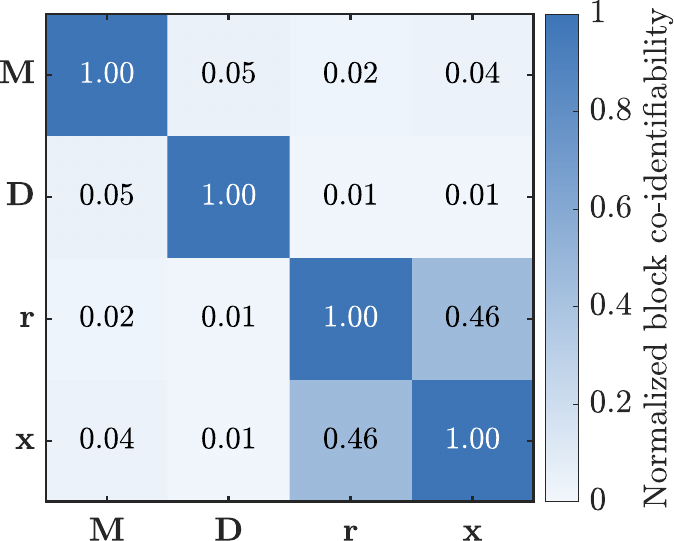} 
\caption{Normalized co-identifiability matrix \(I_{XY}\), showing the strength of cross-block coupling in \(\boldsymbol{\theta}=[\boldsymbol{M}^\top,\boldsymbol{D}^\top,\boldsymbol{r}^\top,\boldsymbol{x}^\top]^\top\).}
\label{fig:co_identifiability} 
\end{figure}

Fig.~\ref{fig:co_identifiability} displays the normalized co-identifiability matrix $I_{XY}$~\eqref{eq:coid} for blocks $X,Y\in\{\boldsymbol{M},\boldsymbol{D},\boldsymbol{r},\boldsymbol{x}\}$, computed from the local Gauss--Newton curvature at the posterior center. Several features are noteworthy. First, inertia $\boldsymbol{M}$ and damping $\boldsymbol{D}$ are only weakly coupled to the network parameters ($I_{M,r}\approx I_{M,x}\approx 0.02$--$0.04$, $I_{D,r}\approx I_{D,x}\approx 0.01$), confirming that generator dynamics are primarily identified from the early-time frequency response regardless of network uncertainty. Second, within the generator block, $I_{M,D}\approx 0.05$, indicating that the inertia-dominated frequency swing and the damping-dominated oscillation decay are distinguishable in the time-segmented likelihood. Third, the network parameters $\boldsymbol{r}$ and $\boldsymbol{x}$ share substantial co-information ($I_{r,x}\approx 0.46$), consistent with the fact that both enter the series admittance nonlinearly through $g^s=r/(r^2+x^2)$ and $b^s=-x/(r^2+x^2)$, creating a coupled sensitivity structure in the power flow manifold. This coupling motivates the separate resistance and reactance blocks in the Gibbs sampler (Section~\ref{subsec:blocked}), which allows each network-parameter class to be updated with the other held fixed, mitigating the confounding.

\begin{figure*}[t]
\centering
\includegraphics[width=\linewidth]{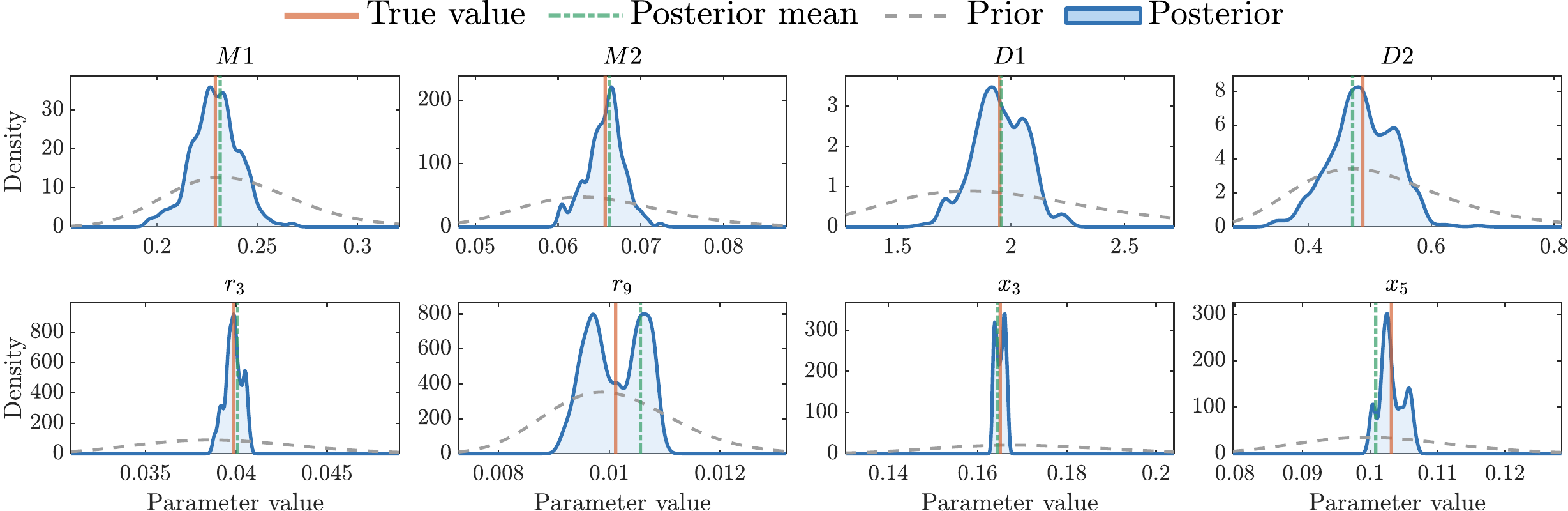}
\caption{Prior-to-posterior update for representative parameters under the joint Bayesian model. Posterior marginals contract around the true values while remaining consistent with the physics-informed priors.
\label{fig:posteriors}}
\end{figure*}

\begin{figure}[t]
\centering
\includegraphics[width=\linewidth]{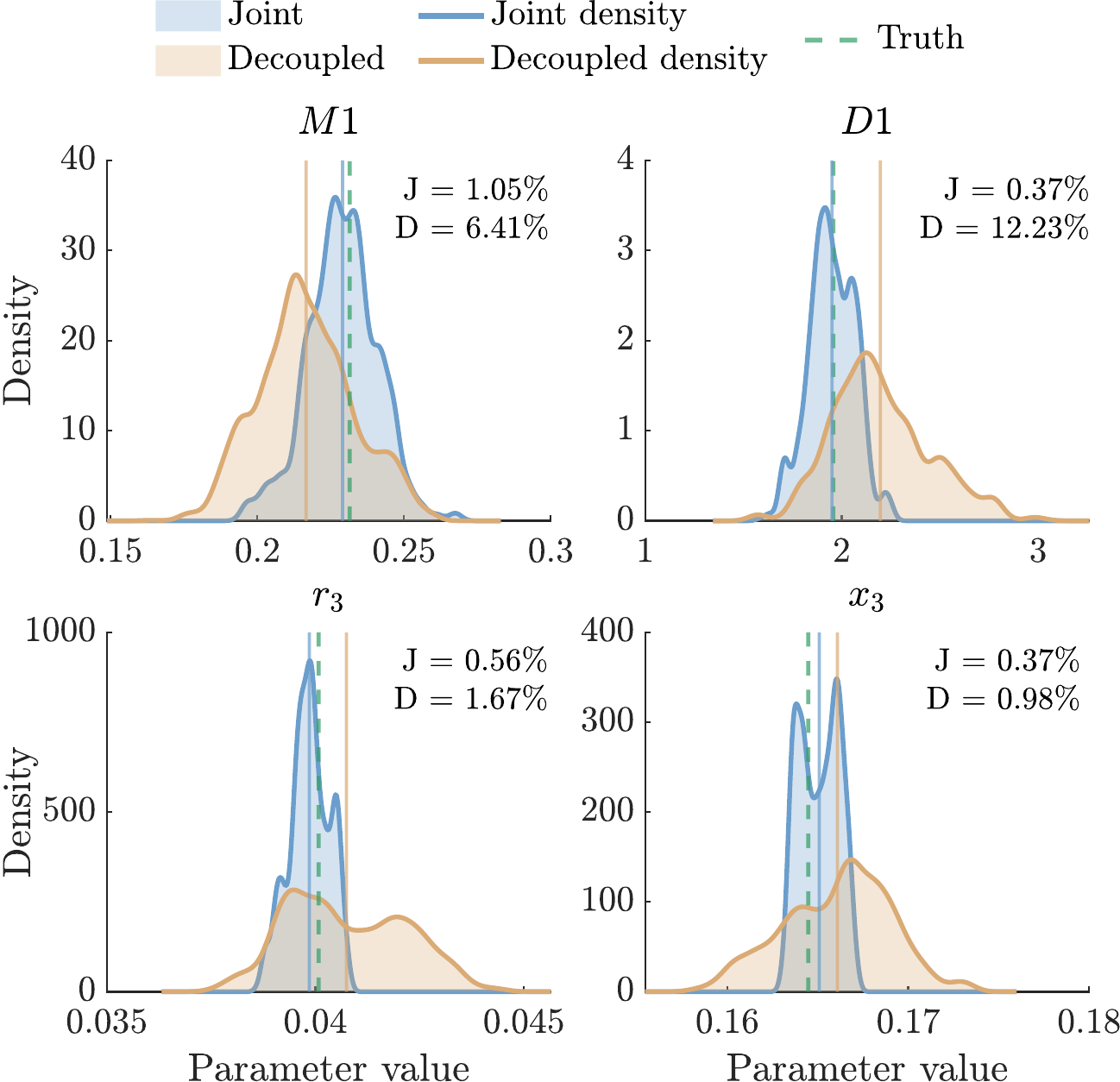}
\caption{Joint and decoupled posterior marginals for selected entries of \((\boldsymbol{M},\boldsymbol{D},\boldsymbol{r},\boldsymbol{x})\). The joint formulation yields better-centered posteriors, highlighting the effect of the DAE coupling encoded in \eqref{eq:dae-compact}--\eqref{eq:pfcons}.}
\label{fig:joint_decoupled}
\end{figure}

\begin{table*}[http]
\centering
\caption{Computational summary for the joint Bayesian estimator.}
\label{tab:main_cost_summary}
\resizebox{\textwidth}{!}{%
\begin{tabular}{lccccccc}
\hline
Run & Iterations & Burn-in / kept & Time [min] & Exact solves & Exact-solve reduction & 95\% CI coverage & Overall acceptance \\
\hline
Joint, reduced-budget & 500 & 300 / 200 & 7.2 & 247 & 50.6\% & 12/21 & 42.0\% \\
Joint, full-budget    & 7000 & 3000 / 2000 & 30.4 & 2093 & 70.1\% & 21/21 & 23.8\% \\
\hline
\end{tabular}}
\end{table*}

\begin{table*}[http]
\centering
\caption{Ablation and comparison summary}
\label{tab:ablation_summary}
\resizebox{\textwidth}{!}{%
\begin{tabular}{lcccccc}
\hline
Variant & Budget & Exact solves & Time [min] & Main quantitative result & Interpretation \\
\hline
Blocked + DA & 500 iter & 247 & 7.2 & Ref. short-budget sampler; $\boldsymbol{D}$ error $=8.4/19.6\%$ & Best cost-accuracy trade-off \\
Blocked + exact-only & 500 iter & 500 & -- & Median posterior-mean shift vs.\ DA $=0.589\%$ & DA preserves posterior, reduces exact solves by 50.6\% \\
Full-block + DA & 500 iter & 259 & 7.6 & $\boldsymbol{D}$ error $=14.3/26.2\%$ & Full-block proposals less effective than blocked updates \\
Joint inference & long run & 2093 & 30.4 & $\boldsymbol{M}=0.7/1.1\%$, $\boldsymbol{D}=1.7/3.5\%$ & Best recovery under the coupled DAE model \\
Decoupled inference & long run & -- & -- & $\boldsymbol{M}=9.6/16.7\%$, $\boldsymbol{D}=12.6/23.9\%$ & Fixing one parameter class biases the other \\
\hline
\end{tabular}}

\end{table*}

\subsection{Joint Estimation Results}

Fig.~\ref{fig:posteriors} overlays prior (dashed) and posterior (solid, shaded) marginal densities for eight representative parameters (two inertias ($M_1, M_2$), two dampings ($D_1, D_2$), two resistances ($r_3, r_9$), and two reactances ($x_3, x_5$)) together with the true value (solid vertical line) and posterior mean (dashed vertical line). In all cases the posterior contracts sharply relative to the weakly informative prior, and the true value falls within the high-density region. The inertia and damping marginals are smooth and approximately Gaussian, reflecting the strong information content of frequency channels for these parameters. The resistance and reactance marginals are tighter still (note the compressed horizontal axes), consistent with the moderate voltage-channel inflation ($\kappa_{\mathrm{volt}}=5$) that preserves network-parameter identifiability while preventing pathological posterior sharpness.
Across all 21 parameters, the posterior-mean relative errors are: $\boldsymbol{M}$ mean/max $=[0.7\%,\,1.1\%]$, $\boldsymbol{D}$ mean/max $=[1.7\%,\,3.5\%]$, $\boldsymbol{r}$ mean/max $=[2.2\%,\,4.2\%]$, $\boldsymbol{x}$ mean/max $=[0.5\%,\,2.3\%]$. All true values lie within the 95\% credible intervals. The MCMC diagnostics confirm healthy sampling: the overall acceptance rate is 23.8\%, the Stage~1 acceptance rate is 29.9\%, the Stage~2 conditional acceptance is 79.7\%, and the power-flow rejection rate at Stage~0 is negligible. As a trajectory-level check, re-simulating the DAE at the posterior representative parameter vector yields a clean-trajectory mismatch of $\mathrm{MSE}=1.04\times10^{-8}$, with residual RMS values $[\mathrm{all},\,\mathrm{freq},\,\mathrm{volt}] = [1.087\times10^{-4},\,1.330\times10^{-5},\,1.196\times10^{-4}]$.

\subsection{Computational Cost, Sampler Ablation, and Scalability}\label{subsec:cost_ablation}

The dominant cost of the proposed method is repeated forward evaluation of the DAE model~\eqref{eq:dae-compact} inside the posterior~\eqref{eq:logpost_eta}. For the IEEE 9-bus case ($n_\theta=21$, $n_E=4$, $T=10$\,s), the reported full run uses 7000 MCMC iterations (3000 burn-in, 2000 retained samples with thinning factor 2). The total wall-clock time is 30.4\,min on a single-core laptop, of which approximately 4\,min are spent on stagewise initialization (Section~\ref{subsec:init}), 1\,min on local curvature estimation for proposal preconditioning, and 25\,min on MCMC sampling. In this run, the multifidelity delayed-acceptance sampler required 6992 coarse DAE evaluations and only 2093 exact DAE evaluations, corresponding to a 70.1\% reduction in exact solves relative to an exact-only sampler of the same length. The measured acceptance statistics were 29.9\% at Stage~1, 79.7\% at Stage~2 conditional on Stage~1 acceptance, and 23.8\% overall, while Stage~0 power-flow rejection remained negligible. We note that log-posterior traces and acceptance statistics indicate that the sampler has largely stabilized by about 2000 iterations, so the full 7000-iteration run should be interpreted as an accuracy-focused setting rather than strict computational requirement. Table~\ref{tab:main_cost_summary} summarizes the main computational results.

These numbers clarify the role of the delayed-acceptance construction in Section~\ref{subsec:delayed}. The coarse model used in Stage~1 is not a different physics model, it is the same DAE solved on a sparser time grid and with relaxed tolerances. Its purpose is purely computational: proposals that are already unlikely under a highly correlated cheap approximation are filtered before the expensive exact evaluation is performed. The Metropolis correction~\eqref{eq:alpha2} ensures that the chain still targets the exact posterior~\eqref{eq:posterior_alpha}, so the coarse stage changes efficiency, not the stationary distribution.
We also performed a matched-budget sampler ablation to isolate the contribution of each sampling component, see Tab.~\ref{tab:ablation_summary}. First, comparing blocked Gibbs proposals (Section~\ref{subsec:blocked}) against a full-block proposal under the same 500-iteration short budget shows that the blocked scheme is materially more reliable: the blocked sampler achieves substantially better parameter accuracy, especially for the damping parameters, whereas the full-block proposal exhibits noticeably poorer centering and broader posterior error. This is consistent with the blockwise co-identifiability structure in Fig.~\ref{fig:co_identifiability}: updating all 21 parameters simultaneously creates a harder sampling geometry than cycling through dynamics, resistance, and reactance subspaces separately.
Second, to verify that delayed acceptance does not distort inference, we compared the blocked delayed-acceptance sampler against a blocked exact-only sampler at the same short budget. The delayed-acceptance variant used only 247 exact solves versus 500 for the exact-only sampler, i.e., a 50.6\% reduction in exact DAE evaluations, while the median relative shift in posterior means between the two samplers was only 0.589\%. This confirms that the  Stage~1 filter delivers a computational saving without materially altering posterior summaries.

In terms of scalability, the bottleneck is the cost of each forward DAE solve rather than the algebra of the MCMC accept-reject step itself. The present framework is therefore best viewed as an offline calibration tool for small to medium systems. For substantially larger networks, further acceleration would require sparse linear algebra, parallel forward solves, or more aggressive surrogate modeling at Stage~1.

\subsection{Joint vs.\ Decoupled Estimation}

To demonstrate why joint estimation matters, we run a decoupled baseline that estimates $(\boldsymbol{M},\boldsymbol{D})$ with $(\boldsymbol{r},\boldsymbol{x})$ fixed at nominal, and separately estimates $(\boldsymbol{r},\boldsymbol{x})$ with $(\boldsymbol{M},\boldsymbol{D})$ fixed at nominal. Both use the same MCMC machinery, noise model, and priors as the joint run.
Fig.~\ref{fig:joint_decoupled} compares joint and decoupled posterior marginals for one parameter from each class ($M_1$, $D_1$, $r_3$, $x_3$). The effect of decoupling is striking. For $D_1$, the decoupled posterior mean is biased by 12.2\% from the truth (vs.\ 0.4\% for joint), and the posterior is noticeably shifted and broadened. For $M_1$, the decoupled error is 6.4\% (vs.\ 1.1\% joint). The mechanism is precisely the DAE coupling identified in Section~\ref{subsec:coident}: when network parameters are fixed at incorrect nominal values, the optimizer compensates by distorting the generator parameters to absorb the network-induced measurement mismatch. For the network parameters $r_3$ and $x_3$, decoupled errors are 1.7\% and 1.0\% (vs.\ 0.6\% and 0.4\% joint)---smaller but still systematically biased, because fixing dynamics at nominal introduces a complementary error in the voltage-channel fit. The joint formulation avoids this cross-contamination by sampling all parameter classes simultaneously, allowing the posterior to correctly attribute measurement variations to their physical sources through the DAE structure~\eqref{eq:dae-compact}--\eqref{eq:pfcons}.

\subsection{Extension to Bigger Networks}

To assess behavior beyond the 9-bus benchmark, we ran a short (1000 iterations) joint estimation experiment on the 39-bus system. This is substantially harder than the setups used in many Bayesian power system PE studies as it requires jointly estimating 108 parameters. Even with only limited tuning beyond the 9-bus configuration, the posterior-mean errors remained modest and the reconstructed trajectories matched the clean response. At the same time, the posterior intervals remained under-dispersed, with only 14 of 108 true values lying inside the nominal 95\% credible intervals. We therefore view this result as encouraging proof of concept for larger coupled systems, but not yet as a fully tuned large-scale validation. Most parameters were recovered with low relative error, while the main remaining issue in this larger case was uncertainty calibration rather than point-estimate accuracy, which can be significantly improved with careful tuning. Tab.~\ref{tab:case39_short} summarizes the main quantitative results for the 39-bus experiment.
\begin{table}[t]
\centering
\caption{Summary of the short 39-bus joint estimation experiment.}
\label{tab:case39_short}
\small
\resizebox{0.8\columnwidth}{!}{%
\begin{tabular}{lc}
\toprule
Metric & Value \\
\midrule
Unknown parameters & 108 \\
Mean rel. error in $\boldsymbol{M}$ & 3.32\% \\
Mean rel. error in $\boldsymbol{D}$ & 3.98\% \\
Mean rel. error in $\boldsymbol{r}$ & 4.82\% \\
Mean rel. error in $\boldsymbol{x}$ & 2.45\% \\
95\% CI coverage & 14/108 \\
Trajectory mismatch (MSE) & $9.36\times10^{-8}$ \\
Wall-clock time & 65.8 min \\
\bottomrule
\end{tabular}}
\end{table}

\section{Conclusion, Limitations, and Future Work}
\label{sec:conc}
This paper developed a Bayesian framework for joint estimation of generator parameters $(\boldsymbol{M},\boldsymbol{D})$ and network branch parameters $(\boldsymbol{r},\boldsymbol{x})$ in multimachine power-system DAE models using PMU voltage and frequency measurements. The results show that joint estimation is feasible and that decoupling the two parameter classes can introduce noticeable bias.
The study also has clear limitations. The validation is limited to a small synthetic test system, so the results should be interpreted as a proof of concept rather than evidence of broad scalability. Because both data generation and inference use the same DAE model class, the study remains subject to the usual inverse-crime limitation, although measurement noise and parameter perturbations partially reduce this effect. The method is also computationally expensive and is intended for offline calibration, not real-time use.
Future work should therefore focus on validation under stronger model mismatch, larger networks, and real PMU data.

\bibliographystyle{IEEEtran}
\bibliography{refs}
\end{document}